\documentclass{aa}  

\usepackage{graphicx}
\usepackage{txfonts}
\usepackage{natbib}
\usepackage{xcolor}
\usepackage{color, soul}
\usepackage{enumerate}
\usepackage{booktabs}
\usepackage{appendix}
\usepackage{float}
\usepackage{placeins}
\usepackage{bm}
\bibpunct{(}{)}{;}{a}{}{,}
\begin{document} 

\title{\textcolor{black}{Europium as a lodestar: diagnosis of radiogenic heat production in terrestrial exoplanets}}
\subtitle{\textcolor{black}{Spectroscopic determination of Eu abundances in $\alpha$ Centauri AB}   	
\thanks{Based on observations collected at the La Silla Observatory, ESO (Chile) with the HARPS spectrograph.}
}
   \titlerunning{Europium as a radiogenic heat-power diagnostic}
  \authorrunning{H. S. Wang et al.}
   \author{\textcolor{black}{H. S. Wang\inst{1},  
   	      T. Morel\inst{2},
   	      S. P. Quanz\inst{1},
   	      and
   	      S. J. Mojzsis\inst{3,4}          
           }}

   \institute{
   	Institute for Particle Physics and Astrophysics, ETH Z\"{u}rich, Wolfgang-Pauli-Strasse 27, 8093 Z\"{u}rich, Switzerland\\
   	\email{haiwang@phys.ethz.ch; \textcolor{black}{sascha.quanz@phys.ethz.ch}}
   	\and
   	Space sciences, Technologies and Astrophysics Research (STAR) Institute, Universit\'e de Li\`ege, Quartier Agora, All\'ee du 6 Ao\^ut 19c, B\^at. B5C, B4000-Li\`ege, Belgium\\
    \email{tmorel@uliege.be}             
    \and
    \textcolor{black}{Department of Geological Sciences, University of Colorado, UCB 399, 2200 Colorado Avenue, Boulder, CO 80309-0399, USA
    \textcolor{black}{\email{stephen.mojzsis@colorado.edu}}}
    \and
    \textcolor{black}{Institute for Geological and Geochemical Research, Research Centre for Astronomy and Earth Sciences, Hungarian Academy of Sciences, H-1112 Budapest, Hungary}}    
      
 \date{Received 09 May 2020 / Accepted 12 October 2020}

 \abstract
 {Long-lived radioactive \textcolor{black}{nuclides}, such as \textcolor{black}{$^{40}$K, $^{232}$Th, $^{235}$U and $^{238}$U,} contribute 
 to persistent heat production in the mantle of \textcolor{black}{terrestrial-type} planets. As refractory elements, the concentrations of Th and U in a terrestrial exoplanet are \textcolor{black}{implicitly} reflected in the photospheric abundances in the stellar host. However, a robust determination of these stellar abundances is difficult in practice owing to the \textcolor{black}{general} paucity and weakness of the relevant spectral features.}
 {We draw attention to the refractory, $r-$process element europium, which \textcolor{black}{may} be used as a convenient and \textcolor{black}{practical} proxy for \textcolor{black}{the population analysis of} radiogenic heating in exoplanetary systems.} 
 {As a \textcolor{black}{case study}, we present a determination of \textcolor{black}{Eu} abundances in the photospheres of $\alpha$ Cen A and B \textcolor{black}{with high-resolution HARPS spectra and a strict line-by-line differential \textcolor{black}{analysis}. To first order, the measured Eu abundances can be converted to the abundances of $^{232}$Th, $^{235}$U and $^{238}$U with observational constraints while the abundance of $^{40}$K is approximated independently with a Galactic chemical evolution model.}}
 {\textcolor{black}{Our determination} shows that europium is depleted with respect to iron by $\sim$ 0.1 dex \textcolor{black}{and to silicon by $\sim$ 0.15 dex} compared to solar in both binary components. \textcolor{black}{The} \textcolor{black}{loci} \textcolor{black}{of $\alpha$ Cen AB at the low-ends of both [Eu/Fe] and [Eu/Si] distributions of a large sample of FGK stars further suggest significantly lower potential of radiogenic heat production in any putative terrestrial-like planet (i.e. $\alpha$-Cen-Earth) in this system, compared to that in rocky planets (including our own Earth) formed around the majority of these Sun-like stars. Based on our calculations of the radionuclide concentrations in the mantle and assuming the mantle mass} \textcolor{black}{to be the} \textcolor{black}{same as that of our Earth, we find that the radiogenic heat budget in an $\alpha$-Cen-Earth is $73.4^{+8.3}_{-6.9}$ TW upon its formation and $8.8^{+1.7}_{-1.3}$ TW \textcolor{black}{at} the present day, respectively $23\pm5$\% and $54\pm5$\% lower than that in the Hadean Earth ($94.9\pm5.5$ TW) and in the modern Earth ($19.0\pm1.1$ TW).}}  
 {\textcolor{black}{As a consequence, mantle convection in an $\alpha$-Cen-Earth} 
 \textcolor{black}{is expected to be overall} \textcolor{black}{weaker than that of Earth (assuming other conditions are the same) and thus such a planet would be less geologically active, suppressing its} \textcolor{black}{long-term potential to recycle its crust and volatiles.} \textcolor{black}{With Eu abundances \textcolor{black}{being} available for a large sample of Sun-like stars, the proposed approach can extend our ability to make predications \textcolor{black}{about} the nature of other rocky worlds that can be tested by future observations.}}
 	


\maketitle

\section{Introduction}\label{sect_intro}

 A major goal of modern astronomy is to \textcolor{black}{better define} the nature of exoplanets and understand planet formation and evolution, as well as life prospects, in a cosmic perspective. To this end, communities in Earth sciences, exoplanet science, and stellar astrophysics have been increasingly joining forces, through \textcolor{black}{existing and future}, ground- and space-based astronomical infrastructures (e.g. Gaia, VLT, Kepler, TESS, ELT, JWST, PLATO) and/or \textcolor{black}{collaboration networks} (e.g. NExSS, PlanetS, GALAH). 

 Through spectroscopic observations of their photospheres, one can decipher 
 \textcolor{black}{the elemental compositions of stars}, which in turn \textcolor{black}{yield unique insights into the formation and bulk compositions} of the planets formed around them \citep{Bond2010, Pagano2014, Wang2019a, Doyle2019, Liu2020}. The \textcolor{black}{resulting data can be used to infer gross geodynamical properties} (including interior, surface, atmosphere, as well as habitability) of \textcolor{black}{terrestrial-like} exoplanets \citep{Frank2014, Noack2017, Hinkel2018, Wang2019b, Shahar2019}.

 \textcolor{black}{Radiogenic heat generated by the decay of the} \textcolor{black}{long-lived radionuclides ($^{40}$K, $^{232}$Th, $^{235}$U and $^{238}$U) contributes a time-\textcolor{black}{dependent} but significant proportion of the Earth's internal heat \citep[e.g.][]{Gando2011, Lenardic2011, Frank2014, Nimmo2015}\footnote{\textcolor{black}{The heat-producing short-lived nuclides $^{26}$Al and $^{60}$Fe are important heating sources in shaping the composition of planetesimals in the early solar system \citep{Lichtenberg2016} but become effectively extinct after 3 Myrs and essentially have no contribution to heat production in \textcolor{black}{already} formed planets \citep{Frank2014}.}}, which powers mantle convection leading to persistent plate tectonics, plume activity and other forms of volcanism that eventually made our planet habitable \citep{Sleep2007,lugaro18, Lingam2020, Seales2020}.}  Therefore, knowing the abundances of these long-lived, heat-producing isotopes in other rocky planets is critical \textcolor{black}{to the assessment (to first order) of the geological activity of these planets.} 
 
 \textcolor{black}{The other main source of planetary internal heat is the dynamical/gravitational energy, inherited from planet formation and core-mantle segregation\textcolor{black}{; this} declines with time \textcolor{black}{owing to} secular cooling \citep{Stevenson2003, Lyubetskaya2007}. (Radiogenic heating also declines with time but in different ways, i.e. following the exponent decay of radionuclides.) \textcolor{black}{An expression} of the relative importance between radiogenic heating and the gravitational energy is \textcolor{black}{dubbed} the Urey ratio/number, which is defined as the ratio of the instantaneous radiogenic heat production to the total surface heat flow of the planet at that time\textcolor{black}{; proposed values range} from approximately 0.3 to 0.9 for the case of the Earth \citep{Schubert2001}\footnote{\textcolor{black}{Another similar concept is called "the convective Urey ratio" \citep{Korenaga2008}, which refers to the mantle contribution alone. However, we do not distinguish it from the (total) Urey number we have adopted here since that would require the crustal information, which is far \textcolor{black}{from being ascertained with} our current observations of exoplanets.}}. \textcolor{black}{Yet, placing direct constraints} on the Urey number} 
 \textcolor{black}{for exoplanets is \textcolor{black}{not possible} due to the paucity of information related to the \textcolor{black}{multitude of possible} disc environments, planet formation histories and evolution scenarios.} 
 \textcolor{black}{With these important caveats, it makes sense to start with} \textcolor{black}{the host stellar abundances} 
\textcolor{black}{to make \textcolor{black}{preliminary} inferences about rocky worlds around other stars \textcolor{black}{\citep{Santos2017, Doyle2019, Wang2019b, Liu2020}} that can be verified (or refuted) as new observational techniques come online.}
	
 \textcolor{black}{\textcolor{black}{On top of this}, challenges also exist in spectroscopically determining the abundances of \textcolor{black}{the principal} long-lived radioactive elements in planet-hosting stars} \citep[e.g.][]{Unterborn2015,del_peloso05,botelho19}.\footnote{For instance, the Th abundance is commonly determined through modelling of a single line (\ion{Th}{ii} $\lambda$4019.1) that is heavily blended with stronger features of other elements (e.g. Fe \textcolor{black}{and Ni}) and very sensitive to continuum placement.} As a result, selecting other elements that can act as proxies \textcolor{black}{and are easier to measure is a convenient and practical approach, especially for population analysis of potential rocky worlds around other stars.} 
 
 \textcolor{black}{Owing to the fact that} U and Th are pure neutron-capture (rapid-) $r$-process elements \citep{Simmerer2004, Bisterzo2014}, the surrogates should also be produced through this nucleosynthesis channel as much as possible. Based on this criterion, Ir and Eu -- \textcolor{black}{98.4 and 94.0\%} contributions by $r$-process, respectively \textcolor{black}{\citep{Bisterzo2014}} -- are most suitable proxy candidates for U and Th. Because the Eu abundances are considerably easier to determine in extrasolar systems, we regard it as the appropriate choice here \citep[as supported by][]{Yong2008}. There are also other proxies for long-lived radionuclides discussed in the literature, e.g., Hf, Bi and Tl \citep{Sneden2008, Wilford2011}. However, Eu must clearly be preferred given 
 \textcolor{black}{the large sets} of Eu abundances for FGK dwarfs \citep[e.g.][]{Pagano2014,Delgado2017,battistini16,mishenina16,guiglion18}\textcolor{black}{, which overwhelmingly exceed} what is available for \textcolor{black}{other proxy candidates} \citep[see, e.g.,][]{hinkel19}. \textcolor{black}{The abundance measurements of the latter proxies} \textcolor{black}{are either restricted to stars with very peculiar chemical patterns \citep[e.g.][]{roederer18} or not possible at all\textcolor{black}{, as illustrated by} the absence of their lines in the solar photospheric spectrum \citep[e.g. Bi and Tl;][]{grevesse15}. In addition, because of the high $r$-process contribution to Eu, it has been used extensively together with \textcolor{black}{a typical $s$-process element (e.g. Ba)} as a chemical clock in nucleosynthesis as well as to assess the $r$-process enrichment in galaxy chemical evolution histories \citep{mashonkina00, Jacobson2013, Ji2016, Bisterzo2016, skuladottir19}.}  

 \textcolor{black}{Observations of both Eu and Th abundances in solar analogues with a wide range of ages (0-10 Ga) have shown that} the [Eu/H] and initial [Th/H] abundance ratios evolve in lockstep: [Th/Eu] is solar within 0.04 dex \textcolor{black}{during the Galactic thin disk evolution} \citep{botelho19}. 
 \textcolor{black}{\textcolor{black}{Measurements in meteoritic and in} Galatic halo stars have also shown a nearly constant U/Th production ratio ($0.571^{+0.037}_{-0.031}$; \citealt{Dauphas2005}).}
 In addition, Eu is a refractory element (with a condensation temperature even slightly higher than Fe and Ni; see \citealt{Lodders2003} \textcolor{black}{and \citealt{Wood2019}}), which implies that its abundance in a terrestrial planet is representative and closely related to that of the stellar host. Finally, Eu has stable isotopes and an age correction is thus not necessary to estimate its pristine abundance at the time of planet formation. \textcolor{black}{However, we must note in advance that $^{40}$K cannot be directly/indirectly informed from Eu due to its distinct nucleosynthesis pathways \citep{Clayton2003, Zhang2006} as well as its volatile nature \citep{Wang2019a}. Instead, alternative assumptions (detailed \textcolor{black}{below}) have to be made regarding $^{40}$K.} 

  The recognition that Eu is a suitable reference element for long-lived radionuclides \textcolor{black}{(excluding $^{40}$K)} is by no means new \citep{pagel89}, but one can argue that its usefulness as a proxy for \textcolor{black}{(partial)} internal heat production has not be been fully \textcolor{black}{recognised} by the exoplanet community \textcolor{black}{\citep[e.g.][]{Kite2009}}.  The vast majority of stars potentially hosting terrestrial-like planets actually do not have Eu abundances available \citep[e.g.][]{schuler15}. As demonstrated in the following, \textcolor{black}{Eu abundances} can be \textcolor{black}{spectroscopically} determined with relative ease even in stars that are not necessarily (very) bright, thanks to the moderate strength of the \ion{Eu}{ii} lines in the blue spectral range. Therefore, it is also hoped that our study will \textcolor{black}{motivate} more widespread determinations of the abundance of this element in \textcolor{black}{stars with potentially rocky planets}.

  \textcolor{black}{As a case study}, we \textcolor{black}{turn our attention to} our nearest \textcolor{black}{Sun-like} star systems, $\alpha$ Cen AB, at only $\sim$1.3 pc. Although no planet orbiting either binary component has been confirmed yet \citep[e.g.][]{Dumusque2012, Hatzes2013, Rajpaul2016}, various numerical simulations support the contention that the stability of planetary orbits can persist within the \textcolor{black}{so-called} habitable zones of these two binary stars \citep{Andrade2014, Quarles2016, Quarles2018}. 
  If a small (potentially rocky) planet is discovered in the habitable zone of either $\alpha$ Cen A or B, the \textcolor{black}{intensive} discussion about the \textcolor{black}{nature} of such a planet would be \textcolor{black}{brought to the forefront}. The detailed chemical composition of $\alpha$ Cen AB recently revisited by \citet[][hereafter M18]{morel18} enables such a discussion to be made \textcolor{black}{\textit{quantitatively}}, except for the planetary internal heat budget \textcolor{black}{partially} owing to the lack of abundances of heat-producing elements (U, Th, and K) or their proxies. In the context of exoplanet studies and, in particular, our modelling of the radiogenic heat power in the putative planets orbiting $\alpha$ Cen A or B, we present in this \textcolor{black}{work our spectroscopic analysis and results of Eu abundances of both stars (Sects. \ref{sect_analysis} and \ref{sect_results}),} \textcolor{black}{followed by a} \textcolor{black}{comparison with previous Eu analyses for $\alpha$ Cen A/B (Sect. \ref{sect_disc_comp1}.1) and with Eu abundances of other stars (Sect. 
  \ref{sect_disc_comp2}.2).} \textcolor{black}{We provide an assessment of} \textcolor{black}{radiogenic heat budget for potential, terrestrial planets in the system (Sect. \ref{sect_disc_heat}.3),} \textcolor{black}{\textcolor{black}{together} with an analysis of the limitations to our approach} \textcolor{black}{(Sect. \ref{sect_disc_limit}.4).} \textcolor{black}{Our conclusions are summarised in Sect. \ref{sect_conc}.} 

\textcolor{black}{\section{Analysis}\label{sect_analysis}}

\begin{table*}[!ht] 
	\caption{Changes made to the initial VALD3 line list. LEP is the lower excitation potential.  }
	\label{tab_new_lines} 
	\centering
	\begin{tabular}{llcccl}
		\hline\hline
		Ion                  & $\lambda$ (\AA) & LEP (eV) & \multicolumn{2}{c}{$\log gf$} & Remark\\
		&            &                      & This study & VALD3     & \\
		\hline
		{\bf \ion{Eu}{ii} $\lambda$3907.1} & & & & & \\
		\ion{Fe}{i}          & 3906.9617  & 3.283                & --2.27     & --1.481   & $\log gf$ adjusted\\     
		\ion{Fe}{i}          & 3907.22    & 3.5\tablefootmark{a} & --2.63     & ...       & artificial line   \\
		\ion{Ce}{ii}         & 3907.2876  & 1.107                &  +0.24     &  +0.320   & $\log gf$ adjusted\\      
		\hline                                               
		{\bf \ion{Eu}{ii} $\lambda$4129.7} & & & & & \\
		\ion{Fe}{i}          & 4129.29    & 3.5\tablefootmark{a} & --2.95     & ...       & artificial line   \\
		\ion{Fe}{i}          & 4129.35    & 3.5\tablefootmark{a} & --3.1      & ...       & artificial line   \\
		\ion{Fe}{i}          & 4129.4600  & 3.397                & --1.89     & --1.970   & $\log gf$ adjusted\\       
		\ion{Fe}{i}          & 4129.53    & 3.5\tablefootmark{a} & --3.1      & ...       & artificial line   \\
		\element[][12]{CN}   & 4129.6006  & 0.976                & --0.8      & --0.585   & $\log gf$ adjusted\\
		\ion{Ti}{i}          & 4129.6429  & 2.239                & --1.8      & --1.424   & $\log gf$ adjusted\\       
		\ion{Fe}{i}          & 4129.956   & 3.5\tablefootmark{a} & --2.38     & ...       & artificial line   \\
		\ion{Fe}{i}          & 4130.0366  & 1.557                & --3.54     & --4.034   & $\log gf$ adjusted\\       
		\ion{Cr}{i}          & 4130.0614  & 2.913                & --2.6      & --0.840   & $\log gf$ adjusted\\       
		\hline                                              
		{\bf \ion{Eu}{ii} $\lambda$6645.1} & & & & & \\
		\element[][12]{CN}   & 6644.9153  & 1.066                & --1.825    & --2.008   & $\log gf$ adjusted\\
		\ion{Si}{i}          & 6645.2099  & 6.083                & --2.4      & --3.156   & $\log gf$ adjusted\\       
		\ion{Fe}{i}          & 6645.3645  & 4.386                & --2.95     & --3.622   & $\log gf$ adjusted\\      
		\element[][12]{CN}   & 6645.4038  & 0.956                & --2.087    & --1.887   & $\log gf$ adjusted\\
		\bottomrule 
	\end{tabular}
	\tablefoot{\\
		\tablefoottext{a}{Arbitrary value adopted following \citet{lawler01}.}
	}
\end{table*}

\begin{figure*}[!ht] 
	\centering
	\includegraphics[trim=40 255 30 200,clip,width=0.9\hsize]{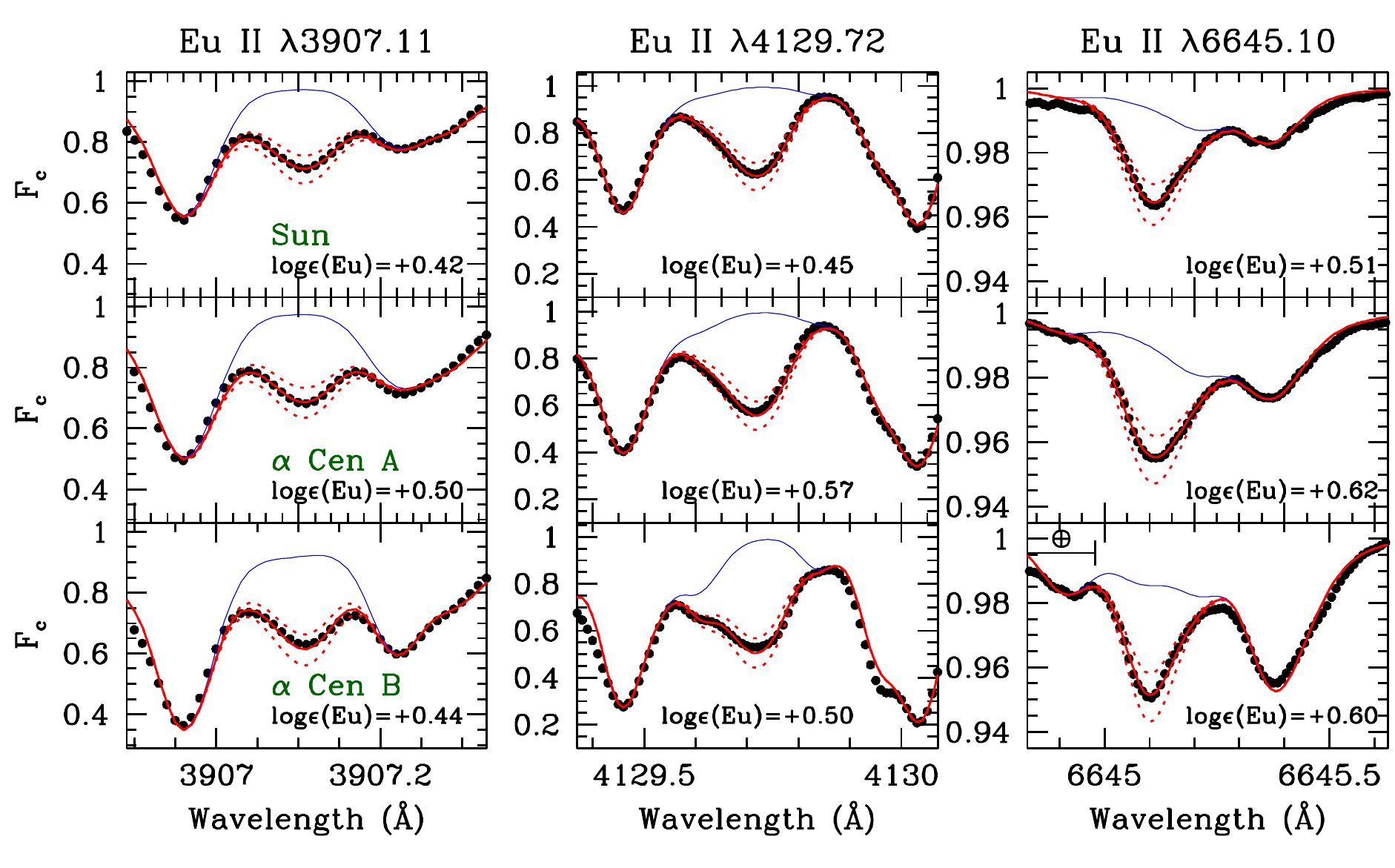}
	\caption{Examples of fits to the \ion{Eu}{ii} spectral features in the Sun ({\it top panels}) and \object{$\alpha$ Cen AB} ({\it middle and bottom panels}) based on the line list of Me14. The solid red line shows the best-fitting synthetic profile, while the two dotted lines show the profiles for an Eu abundance deviating by $\pm$0.1 dex. The blue line shows the profile for no Eu present. The best-fitting abundance is indicated in each case. Note the widely different ordinate scale for \ion{Eu}{ii} $\lambda$6645. The Earth symbol in the bottom right panel marks the location of telluric features (the Sun and \object{$\alpha$ Cen A} are not significantly affected).}
	\label{fig_synthesis}
\end{figure*}

Aiming for a homogeneous spectroscopic analysis of the Eu abundances in $\alpha$ Cen AB, we closely follow the procedures employed in M18. However, two aspects are necessarily different. First, the diagnostic Eu lines are either located in the far blue or \textcolor{black}{otherwise weak}. \textcolor{black}{This means} that the HARPS spectra used by M18, which do not extend below 4800 \AA \ and do not have an extremely high signal-to-noise ratio (S/N), are not suitable. Indeed, M18 could not confidently measure the \ion{Eu}{ii} $\lambda$6645 line with an equivalent width (EW) below 10 m\AA \ in \object{$\alpha$ Cen AB}. Second, the \ion{Eu}{ii} lines are in general strongly blended and broadened by isotopic and hyperfine (HFS) splitting, which calls for an abundance determination relying on spectral synthesis instead of an analysis based on EWs.

\textcolor{black}{For the purpose of this work, our analysis is based on} high-resolution HARPS spectra retrieved from the European Southern Observatory (ESO) archives. \textcolor{black}{These} have a resolving power, $R$, of about 115,000 and cover the spectral range 3780--6910 \AA. For the solar spectrum to be used as reference, we averaged with a weight \textcolor{black}{that depends} on the mean S/N all the available exposures of asteroids with a S/N above 100. Observations of a point-like source \textcolor{black}{are} preferred \citep[e.g.][]{gray00}, while co-adding spectra from various reflecting bodies is not an issue \citep[e.g.][]{bedell14}. For \object{$\alpha$ Cen AB}, we collected all spectra with 375 $<$ S/N $<$ 400. Spectra with a higher S/N were ignored to avoid saturation problems. For \object{$\alpha$ Cen A}, we only considered the numerous spectra obtained over five consecutive nights by \citet{bazot07} for their asteroseismic analysis. After rejecting the spectra with Eu features affected by cosmic rays or telluric features \textcolor{black}{(i.e. the molecular lines of the Earth's atmosphere)}, we ended up with a total of 177, 283, and 284 spectra for the Sun, \object{$\alpha$ Cen A} and \object{$\alpha$ Cen B}, respectively. However, about 30\% of the exposures were discarded for the analysis of \ion{Eu}{ii} $\lambda$6645 in \object{$\alpha$ Cen B} because of fringing patterns in the red. All spectra were corrected from radial-velocity shifts prior to co-adding based on the precise cross-correlation (CCF) data from the instrument reduction pipeline. The mean spectra were normalised to the continuum by fitting low-order Legendre polynomials using standard tasks implemented in the IRAF\footnote{{\tt IRAF} is distributed by the National Optical Astronomy Observatories, operated by the Association of Universities for Research in Astronomy, Inc., under cooperative agreement with the National Science Foundation.} software. To ensure the highest consistency, this procedure was identical for \textcolor{black}{all} three stars.

Our results are based on a line-by-line differential analysis relative to the Sun. We make use of plane-parallel, 1D MARCS model atmospheres \citep{gustafsson08} and the 2017 version of the line-analysis software MOOG originally developed by \citet{sneden73}. M18 carried out the analysis using various line lists taken from the literature. However, only two include Eu features: \citet[][hereafter Me14]{melendez14} considered \ion{Eu}{ii} $\lambda$3819.7, 3907.1, 4129.7 and 6645.1, while the study of \citet[][hereafter Re03]{reddy03} only included \ion{Eu}{ii} $\lambda$6645.1. We do not consider \ion{Eu}{ii} $\lambda$3819.7 any further because this weak line is heavily blended and difficult to model properly, as discussed by \citet{lawler01}. The HFS data are taken from \citet{ivans06} and assume the $^{151}$Eu/$^{153}$Eu isotopic ratio from \citet{chang94}. The study of \citet{ivans06} is an improvement over the reference work of \citet{lawler01}, as it provides updated Eu transition data. About 30 HFS components are taken into account for each line studied. The lines of other elements in the relevant spectral ranges were modelled using data retrieved from the VALD3 atomic database\footnote{http://vald.astro.uu.se/} and assuming the abundances of M18. For the minor species not included in M18, we scaled the abundances according to [Fe/H] because estimates in the literature are either not robust or for the most part simply not available. \textcolor{black}{We note that this assumption has no impact on our main results.} The dissociation energies implemented in MOOG were assumed for the molecular species. To fit the solar spectrum, we adopted projected rotational and macroturbulent velocities of 1.8 and 3.1 km s$^{-1}$, respectively. For \object{$\alpha$ Cen AB}, we assumed the values quoted by \citet{bruntt10}. Instrumental broadening was also taken into account: the value was first adjusted based on a fit of the relatively unblended \ion{Fe}{ii} 4128.7 line \citep[see, e.g.,][]{koch02}, and then scaled as a function of wavelength according to the full width at half-maximum (FWHM) of lines measured in calibration lamps.

Our strategy was first to adjust the oscillator strengths of lines in vicinity of the Eu feature of interest to optimise the quality of the fit in the Sun. The solar mixture adopted \citep{grevesse07} is consistent with that adopted for the computation of the MARCS model atmospheres. For the solar parameters, we adopted an effective temperature, $T_\mathrm{eff}$, of 5777 K, a surface gravity, $\log g$, of 4.44, and a microturbulence, $\xi$, of 1 km s$^{-1}$. Finally, to improve the fit in \object{$\alpha$ Cen AB}, we altered within their uncertainties the abundances of elements with lines significantly affecting the Eu feature. Similarly to a number of previous studies \citep[e.g.][]{del_peloso05,lawler01,peek09}, it was occasionally necessary to include in the line list some artificial \ion{Fe}{i} lines because of unaccounted absorption. The changes to the initial VALD3 line list are summarised in Table \ref{tab_new_lines}. We stress that these slight adjustments lead to a noticeably better fit, but have little impact on the resulting Eu abundances. The modelling of \object{$\alpha$ Cen AB} was performed adopting the stellar parameters ($T_\mathrm{eff}$, $\log g$ and $\xi$) derived by M18 for the relevant line list. As for the other elements studied in M18, we used the ``unconstrained'' results quoted in his table B.2 that are obtained without freezing the surface gravity to the asteroseismic value quoted by \citet{heiter15}. Illustrative examples of the fits are shown in Fig.~\ref{fig_synthesis}.

We determine for the Sun an average Eu abundance, $\langle$$\log \epsilon_{\odot}$(Eu)$\rangle$=+0.46$\pm$0.05, based on 1D model atmospheres and assuming local thermodynamic equilibrium (LTE). Theoretical calculations by \citet{mashonkina00} indicate that the abundances determined from \ion{Eu}{ii} features must be corrected upwards in solar-like stars to account for departures from LTE. The non-LTE corrections for the Sun amount on average to +0.03 dex for \ion{Eu}{ii} $\lambda$4429 and \ion{Eu}{ii} $\lambda$6645. On the other hand, 1D-3D corrections appear to be negligible according to CO$^5$BOLD hydrodynamical simulations \citep{mucciarelli08}. We therefore obtain a corrected solar abundance, $\langle<$$\log \epsilon_\odot$(Eu)$\rangle$=+0.49$\pm$0.05, that is fully compatible with the recommended meteoritic and photospheric values that lie in the range 0.51-0.52 \citep[][and references therein]{grevesse15}. However, given the differential nature of our analysis with respect to the Sun, we emphasise that the exact values of the absolute solar abundances have no bearing on our conclusions.

\section{Results}\label{sect_results}
\begin{table*}[!ht] 
	\caption{Abundance results for \object{$\alpha$ Cen AB} and comparison to values in the literature. Our recommended values are the weighted average (by the inverse variance) of the results obtained using the Me14 and Re03 line lists. For the analysis based on the single \ion{Eu}{ii} feature in Re03 line list, we conservatively assumed $\sigma_\mathrm{int}$ = 0.05 dex.}
	\label{tab_abundances} 
	\hskip -0.4cm
	\begin{tabular}{l|ccc|ccc}
		\hline\hline
		& \multicolumn{3}{c}{$[$Eu/H$]$}                                 & \multicolumn{3}{c}{$[$Eu/Fe$]$}  \\
		& \object{$\alpha$ Cen A} & \object{$\alpha$ Cen B} & A--B & \object{$\alpha$ Cen A} & \object{$\alpha$ Cen B} & A--B \\
		\hline
		{\bf This study} & \multicolumn{3}{c|}{} & \multicolumn{3}{c}{} \\
		Me14 line list   & +0.10$\pm$0.04 (3) & +0.05$\pm$0.06 (3)\tablefootmark{a}   &  +0.05$\pm$0.07 (3)\tablefootmark{a} & --0.12$\pm$0.05 (3) & --0.17$\pm$0.07 (3)\tablefootmark{a} &  +0.05$\pm$0.08 (3)\tablefootmark{a} \\
		Re03 line list   & +0.15$\pm$0.06 (1) & +0.16$\pm$0.08 (1)\tablefootmark{a}   & --0.01$\pm$0.10 (1)\tablefootmark{a} & --0.08$\pm$0.07 (1) & --0.07$\pm$0.09 (1)\tablefootmark{a} & --0.01$\pm$0.11 (1)\tablefootmark{a} \\
		Weighted average & +0.12$\pm$0.04 (3) & +0.09$\pm$0.05 (3)\tablefootmark{a}   &  +0.03$\pm$0.06 (3)\tablefootmark{a} & --0.11$\pm$0.04 (3) & --0.14$\pm$0.06 (3)\tablefootmark{a} &  +0.03$\pm$0.07 (3)\tablefootmark{a} \\
		GCE corrected\tablefootmark{b}  & ... & ...                                   & ...                                  & --0.10$\pm$0.04 (3) & --0.13$\pm$0.06 (3)\tablefootmark{a} &  +0.03$\pm$0.07 (3)\tablefootmark{a} \\
		\hline
		{\bf N97}        & +0.15$\pm$0.05 (1) & +0.14$\pm$0.05 (1)                    & +0.01$\pm$0.08 (1)                   & --0.10$\pm$0.06 (1) & --0.10$\pm$0.06 (1)                  & +0.00$\pm$0.09 (1) \\
		{\bf K02}        & ...                & +0.11$\pm$0.08 (1)                    & ...                                  & ...                 & --0.08$\pm$0.05 (1)                  & ...\\
		{\bf G18}        & ...                & ...                                   &  ...                                 & --0.13$\pm$0.14 (3) & --0.03$\pm$0.03 (3)                  & --0.10$\pm$0.15 (3) \\
		\bottomrule 
	\end{tabular}
	\tablefoot{The number in brackets gives the number of lines used. Keywords for literature studies --- N97: \citet{neuforge97}; K02: \citet{koch02}; G18: \citet{guiglion18}.  Note that the last study made use of archival HARPS spectra \textcolor{black}{that are different from ours}. 
		\tablefoottext{a}{These values are affected by a likely underestimation of the abundances in \object{$\alpha$ Cen B} at the $\sim$0.05 dex level (see Sect.~\ref{sect_results}).}
		\tablefoottext{b}{Computed from the $[$Eu/Fe$]$-age relation of \citet{bedell18} \textcolor{black}{(see Sect. 4.1 for details)}. The uncertainty in the slope of the relation 
			was propagated to the abundances.}
	}
\end{table*}

The \textcolor{black}{abundance analysis} results are summarised in Table \ref{tab_abundances}. Following M18, the random uncertainties are computed by adding in quadrature the line-to-line scatter, $\sigma_\mathrm{int}$, and the uncertainties arising from errors in the stellar parameters. The larger uncertainties for \object{$\alpha$ Cen B} arise from the difficulty in modelling the spectrum of relatively cool stars exhibiting molecular features and the fact that the stellar parameters are much more sensitive to the choice of the iron line list (see M18). 

Tests using Kurucz models indicate that the impact of the choice of the family of 1D model atmospheres is similar for $\alpha$ Cen AB and the Sun, and therefore cancels out to first order through a differential analysis. As discussed in Sect. 
\ref{sect_analysis}, non-LTE effects in the Sun are very small for \ion{Eu}{ii}. Although a detailed quantitative investigation is warranted, we thus do not expect differential corrections to significantly bias our results. On the other hand, \citet{mucciarelli08} predict negligible 3D corrections for \ion{Eu}{ii} $\lambda$6645 for stars with parameters representative of those of \object{$\alpha$ Cen AB}. The accuracy of our abundance results strongly depends on the reliability of the stellar parameters assumed. For both stars, 
our $T_\mathrm{eff}$ is compatible within the uncertainty with the value based on VLTI/PIONIER interferometric measurements \citep{kervella17a}. However, the strength of the \ion{Eu}{ii} lines is particularly sensitive to $\log g$. Although our spectroscopic value for \object{$\alpha$ Cen A} is indistinguishable from the accurate asteroseismic estimate, $\log g$ for \object{$\alpha$ Cen B} appears to be underestimated by 0.09 and 0.23 dex using Re03 and Me14 line lists, respectively (see M18). We regard this bias as the most significant source of systematic error because it likely leads to an underestimation of $[$Eu/H$]$ and $[$Eu/Fe$]$ in this star by $\sim$0.05 dex. 

\section{Discussion}\label{sect_disc}
\textcolor{black}{\subsection{Comparison with previous results of europium abundances in $\alpha$ Cen AB}} \label{sect_disc_comp1}
Our abundances are in good agreement with previous \textcolor{black}{estimates reported} in the literature (Table \ref{tab_abundances}), but our study is generally based on more lines and benefits from higher-quality spectroscopic and HFS data. We do not discuss the results of \citet{allende_prieto04} because there is evidence that their $T_\mathrm{eff}$ scale is too cool (see discussion in M18).

To summarise our \textcolor{black}{abundance analysis} results, we find that europium is depleted with respect to iron in \object{$\alpha$ Cen AB} by $\sim$0.1 dex compared to the Sun and that there is a lack of evidence for a different Eu content in the two components. Our study does not support the claim that the [Eu/Fe] values differ by as much as 0.16 dex between the two stars \citep{hinkel13}, but we note that \citet{allende_prieto04} is their sole literature source for Eu. Our conclusions are still valid if the slight underestimation of the abundances in \object{$\alpha$ Cen B} discussed above is taken into account. Generally speaking, we expect the abundances determined through our differential analysis to be more accurate for \object{$\alpha$ Cen A} because, in view of the similarity with the solar parameters, they are much less sensitive to deficiencies in the  modelling of the atmosphere or non-LTE, 3D and atomic diffusion effects. This caveat affecting the abundances of \object{$\alpha$ Cen B} also applies to other differential studies in the literature \citep[e.g.][]{guiglion18}.

The exact production sites of europium are still uncertain. Cataclysmic events, such as compact binary mergers or core-collapse supernovae, are \textcolor{black}{proposed} to play a role in the nucleosynthesis of Eu, although their relative importance is debated \citep[][and references therein]{skuladottir19}. The [Eu/Fe] ratios are \textcolor{black}{weakly modulated} by Galactic chemical evolution (GCE) effects\textcolor{black}{;} the dependence as a function of stellar age appears to be nearly flat \textcolor{black}{\citep{bedell18}}. Assuming an age for \object{$\alpha$ Cen AB} of 6$\pm$1 Ga (M18\textcolor{black}{, and references therein}) and the linear relations between [Eu/Fe] and age for solar analogues of \citet{bedell18}, we attribute only $\sim$0.01 dex of depletion to GCE (Table \ref{tab_abundances}). Similar GCE effects are also suggested for metal-rich stars \citep[][]{delgado_mena19}.

\textcolor{black}{\subsection{Comparison of europium abundances between $\alpha$ Cen AB and other FGK stars}} \label{sect_disc_comp2}
\begin{figure*}[!ht] 
	\centering
	\includegraphics[trim=4cm 5cm 4cm 5cm, scale=0.6,angle=90]{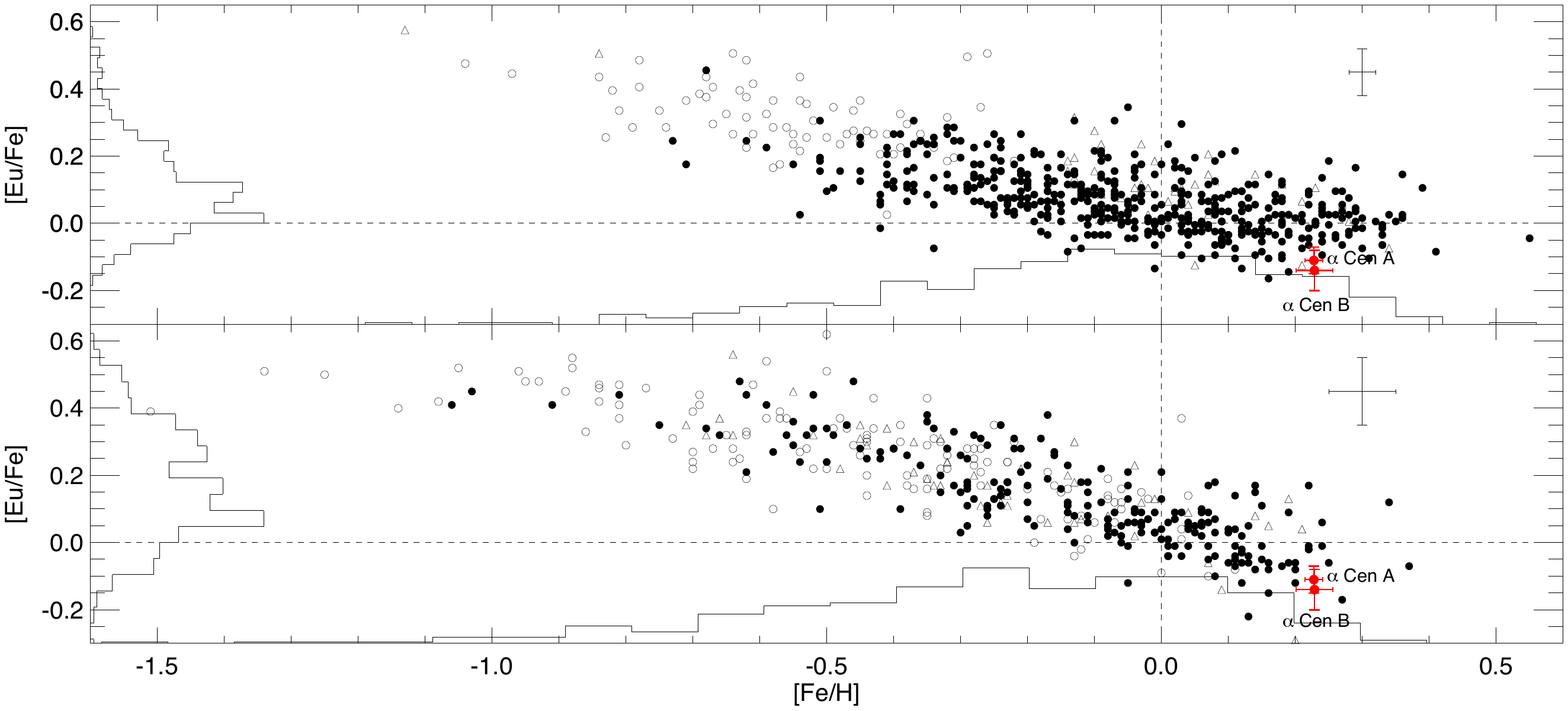}
	\caption{Our abundance results for $\alpha$ Cen A and B \textcolor{black}{overlain} on the [Eu/Fe]-[Fe/H] diagram for large samples of Sun-like stars in the literature. {\it Upper panel}: data for around 570 FGK stars that have available Eu abundances in \citet{Delgado2017, delgado_mena19}. The [Eu/Fe] abundances are taken from \citet{Delgado2017}, but were shifted upwards by 0.016 dex, as recommended by \citet{delgado_mena19}. {\it Lower panel}: same as upper panel, but with data for the 380 FG stars that have available Eu abundances in \cite{battistini16}. The typical uncertainties of the two comparison datasets are shown in the upper right corner of the respective panel. Stars from the thin and thick discs are shown as filled and open circles, respectively, while other stellar populations (i.e. halo, bulge, and uncertain thin-/thick-disc stars) are indicated as open triangles. The histograms on the $x$-axis and $y$-axis show the respective distributions of [Fe/H] and [Eu/Fe] of these FG(K) stars. The horizontal and vertical dashed lines indicate the zero points of the diagrams \textcolor{black}{(solar)}.} 
	\label{fig_comp}
\end{figure*}
In Fig.~\ref{fig_comp} ({\it upper panel}), we compare, as a function of [Fe/H], our [Eu/Fe] abundances of $\alpha$ Cen AB to the values for about 570 FGK dwarfs \citep{delgado_mena19}. Both $\alpha$ Cen A and B appear on the high and low ends of the [Fe/H] and [Eu/Fe] distributions, respectively, demonstrating their unusual position in the [Fe/H]-[Eu/Fe] locus defined by nearby, thin-disc FGK dwarfs. This conclusion may be sensitive to the (unavoidable) presence of zero-point abundance offsets between our study and \citet{delgado_mena19}. However, as seen in Fig.~\ref{fig_comp} ({\it lower panel}), the same picture holds when considering the \textcolor{black}{separate} study of \citet{battistini16}. This indicates that such offsets are unlikely \textcolor{black}{to be} much larger than our abundance uncertainties. Hence, the somewhat peculiar position of $\alpha$ Cen AB on the [Fe/H]-[Eu/Fe] locus \textcolor{black}{is most likely} a reflection of their true nature.

\begin{figure}[h!]
	\centering
	\includegraphics[trim=0.5cm 4cm 0.5cm 4cm, scale=0.4,angle=90]{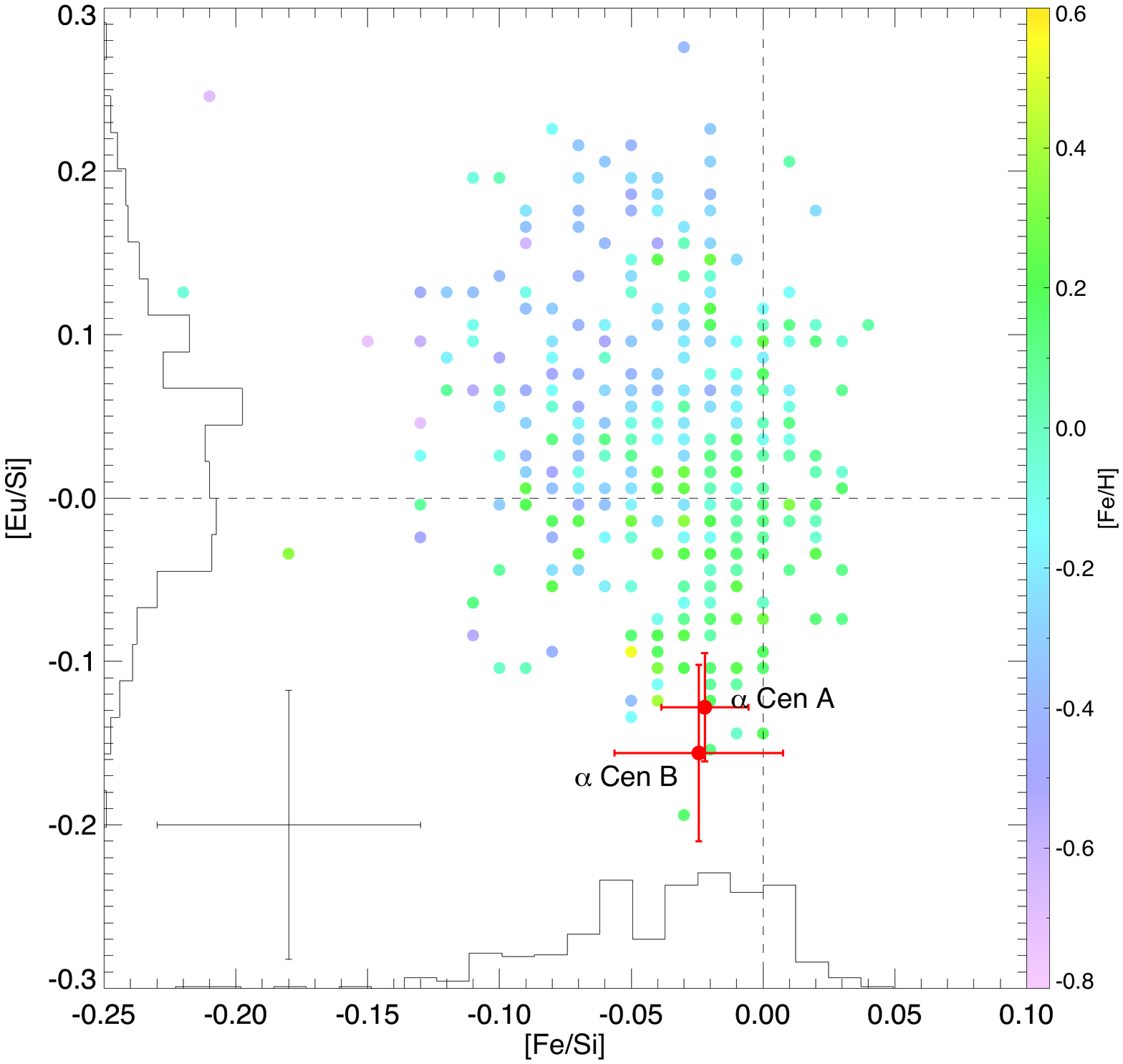}
	\caption{Our abundance results for $\alpha$ Cen A and B \textcolor{black}{overlain} on the [Eu/Si]-[Fe/Si] diagram for the sample of thin-disc stars (colour coded as a function of [Fe/H]) of \citet{delgado_mena19}. Their typical uncertainties are indicated in the bottom left corner.  The histograms on the $x$-axis and $y$-axis show the respective distributions of [Fe/Si] and [Eu/Si] of these thin-disc stars. The horizontal and vertical dashed lines indicate the zero points of the diagram \textcolor{black}{(solar)}.} 
	\label{fig_comp_Si}
\end{figure}

Using another major rock-forming element than Fe as reference \citep[e.g., Si;][]{Unterborn2015} would not qualitatively modify our conclusions given that their abundances are tightly correlated with that of iron in thin-disc stars \citep[e.g.][]{bitsch20}. Indeed, as shown by the [Eu/Si]-[Fe/Si] diagram (Fig. \ref{fig_comp_Si}), $\alpha$ Cen AB still lie in the lower tail of the [Eu/Si] distribution for the sample of thin-disc stars of \citet{delgado_mena19}. However, $\alpha$ Cen AB seem to be not a peculiar, but average case, among these comparison stars in terms of [Fe/Si] that alludes to the first-order, planetary internal structure \citep{Wang2019b}. 

\textcolor{black}{\subsection{Implications \textcolor{black}{for} radiogenic heat budgets in putative $\alpha$-Cen-Earths}}\label{sect_disc_heat}
\begin{table}[!ht] 
\textcolor{black}{
	\caption{Model quantities and results of mantle concentrations and heat output of radionuclides in a putative $\alpha$-Cen-Earth.}
	\label{tab_heat_alpha} 
	\hskip -0.4cm
	\resizebox{\linewidth}{!}{
	\begin{tabular}{cccc}
		\hline\hline
		Constraint & Value & \multicolumn{2}{c}{Reference} \\
		\hline
		$[$Eu/Mg$]$\tablefootmark{a} & $-0.152\pm0.047$ & \multicolumn{2}{c}{\small This work \& M18} \\
		$[$Th/Eu$]$\tablefootmark{b} & $0.014\pm0.045$ & \multicolumn{2}{c}{\small \cite{botelho19}}\\
		$^{238}$U/Th & $0.571^{+0.037}_{-0.031}$ & \multicolumn{2}{c}{\small \cite{Dauphas2005}} \\[2pt]
		$^{235}$U/$^{238}$U\tablefootmark{c} & 24.286/75.712 & \multicolumn{2}{c}{\small \cite{Lodders2009} \& \cite{Frank2014}}\\[2pt]
        $M_{\mathrm{Mantle}}$ (\textcolor{black}{$10^{24}$} kg) & $4.0312\textcolor{black}{\pm}0.0179$ & \multicolumn{2}{c}{\small \cite{Wang2018}}\\
		\hline
		&& \multicolumn{2}{c}{\bf Mantle concentration (ppb)}\\
		Nuclide (X) & X/Mg\tablefootmark{d} &Upon formation\tablefootmark{e} & Present day\tablefootmark{f}\\[2pt]
		$^{232}$Th& $1.95^{+0.64}_{-0.48} \times10^{-7}$ & $43.4^{+14.2}_{-10.7}$ &$32.2^{+10.5}_{-8.0}$\\[2pt]
		$^{235}$U& $3.61^{+1.2}_{-0.91}\times10^{-8}$ & $8.1^{+2.7}_{-2.0}$ &$0.02\pm0.01$\\[2pt]
        $^{238}$U& $1.14^{+0.38}_{-0.29}\times10^{-7}$ &$9.7^{+3.3}_{-2.5}$ & $10.0^{+3.3}_{-2.5}$\\[2pt]
		$^{40}$K & \textcolor{black}{...} & $353.6\pm37.3$\tablefootmark{g} &$12.7\pm1.3$\tablefootmark{g} \\	
		\hline
		&& \multicolumn{2}{c}{\bf Heat (TW)}\\
Nuclide (X) & $h$ (W/kg)\tablefootmark{h} &Upon formation\tablefootmark{i} & Present day\tablefootmark{i}\\[2pt]		
	$^{232}$Th& $2.628\times10^{-5}$ &$4.6^{+1.5}_{-1.1}$&$3.4^{+1.1}_{-0.8}$\\[2pt]
$^{235}$U& $5.6847\times10^{-4}$ &$18.5^{+6.2}_{-4.7}$&$0.05^{+0.02}_{-0.01}$\\[2pt]
$^{238}$U& $9.513\times10^{-5}$ &$9.7^{+3.3}_{-2.5}$& $3.8^{+1.3}_{-1.0}$\\[2pt]
$^{40}$K & $2.847\times10^{-5}$ &$12.7\pm1.3$&$1.5\pm0.2$\\[2pt]
\bf Total & \textcolor{black}{...} &\bm {$73.4^{+8.3}_{-6.9}$}&\bm {$8.8^{+1.7}_{-1.3}$}\\			
		\bottomrule
	\end{tabular}
    }
	\tablefoot{\tablefoottext{a}{$[$Eu/Mg$]$ (dex) is calculated \textcolor{black}{from} [Eu/Fe] - [Mg/Fe], where [Mg/Fe] comes from the weighted average of the results \textcolor{black}{in M18} obtained using the Me14 and Re03 line lists.}\\
		\tablefoottext{b}{100\% of Th is $^{232}$Th.}\\
		\tablefootmark{c}{As explained in text, $^{235}$U/$^{238}$U refers to the solar system value at the time of its formation \citep{Lodders2009} based on the limited variance of the ratio  within 6-9 Ga into the Galactic history \citep{Frank2014}.}\\
   \tablefoottext{d}{X/Mg is a mass ratio. While converting differential values (e.g. [$^{232}$Th/Mg] = [$^{232}$Th/Eu] - [Eu/Mg], in dex) to the mass ratios (e.g. $^{232}$Th/Mg) in \textcolor{black}{a} linear \textcolor{black}{scale}, the \textcolor{black}{reference solar abundances come from \cite{Asplund2009} and the atomic and isotopic masses respectively refer to \cite{Wieser2013} and \cite{Audi1993}}.}\\
   \tablefoottext{e}{See text. $C_0(\alpha\textnormal{-Cen-Earth}) = \frac{\mathrm{X/Mg} (\alpha\textnormal{-Cen-Earth})}{\mathrm{X/Mg} (\textnormal{Earth})}\times C_0(\textnormal{Earth})$, where $C_0$ represents the mantle concentration of an individual radionuclide upon the planet formation; X/Mg ($\alpha$-Cen-Earth) refers to column 2 in panel 2 of this table ($^{40}$K is excluded \textcolor{black}{from} this calculation); X/Mg (Earth) refers to Table \ref{tab_heat_earth} (for \textcolor{black}{the normalisation purpose}, Mg concentration is kept unchanged over geological time).}\\
   \tablefoottext{f}{See text.  $C_t=C_0\times\mathrm{e}^{-t\ln2/T_{1/2}}$, where $t$ represents the system age (the current age of $\alpha$ Cen AB is $6\pm1$ Ga as of M18); $T_{1/2}$ are the half-lives of these radionuclides ($^{232}$Th -- 14.0 Ga; $^{235}$U -- 0.704 Ga; $^{238}$U -- 4.47 Ga; $^{40}$K -- 1.25 Ga; \citealt{Turcotte2002}).}\\
   \tablefoottext{g}{See text. Concentration of $^{40}$K upon planet formation, $C_0(^{40}\mathrm{K})$, is supplemented by the starting mantle concentration of $^{40}$K in the Earth upon its formation, according to the GCE model for cosmochemically Earth-like planets \citep{Frank2014}; the present-day concentration of $^{40}$K is calculated through the radioactive decay process \textcolor{black}{following} the equation in footnote $(f)$.}\\
   \tablefoottext{h}{$h$ (the rate of heating per unit mass of a radionuclide) refers to \cite{Dye2012}.}\\
   \tablefoottext{i}{Calculated by multiplying the corresponding concentrations (upon planet formation and present day) with the heat generation rates, $h$, of individual radionuclides.}\\
	}
}
\end{table}

\begin{figure}[h!]
	\centering
	\includegraphics[trim=0cm 2cm 0cm 2cm, scale=0.45,angle=0]{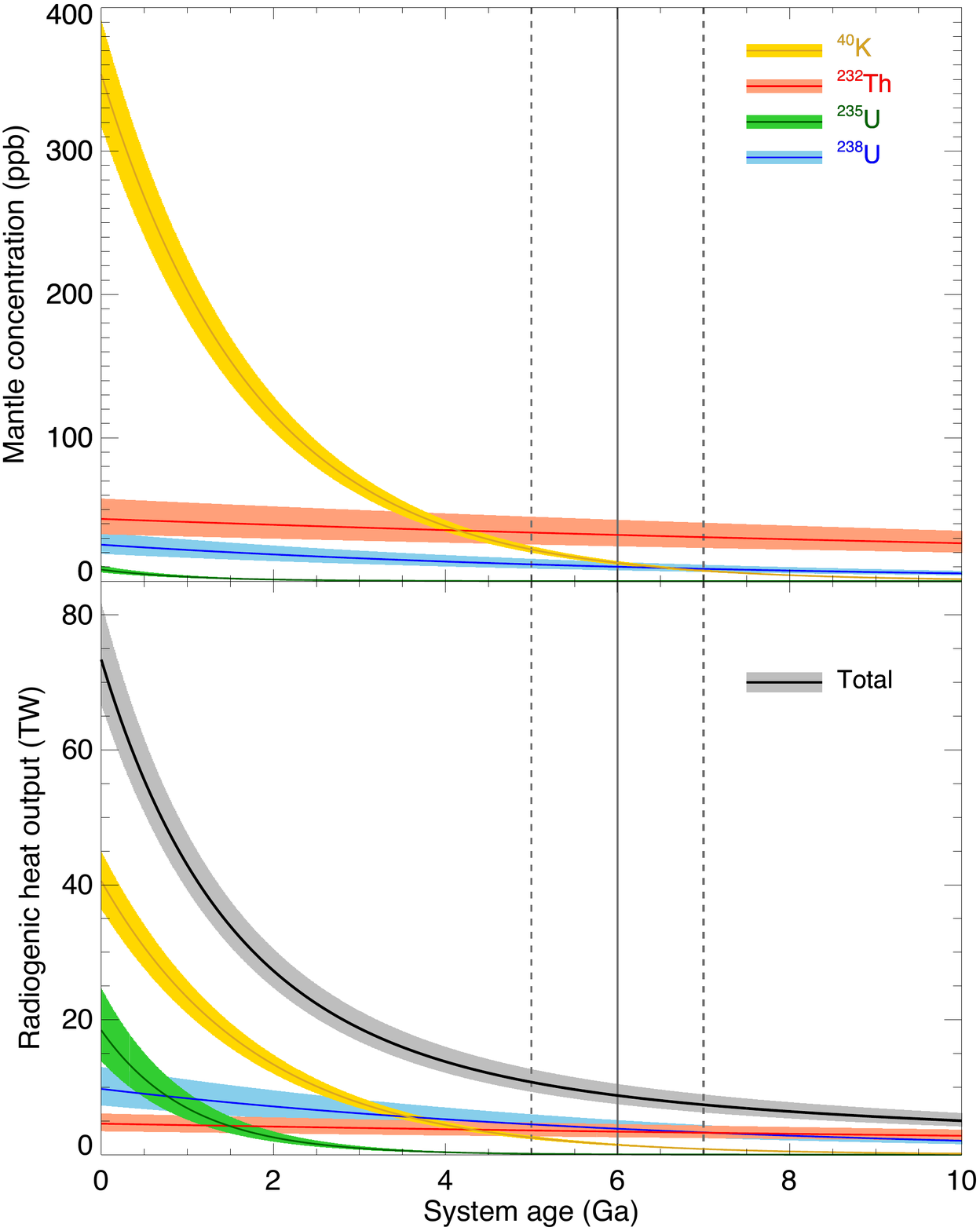}
	\textcolor{black}{
	\caption{Estimated mantle concentrations (upper panel) and radiogenic heat output (lower panel) of $^{40}$K,$^{232}$Th, $^{235}$U, and $^{238}$U in an $\alpha$-Cen-Earth as a function of the system age. The wedge around each curve represents the uncertainties of the estimates over the time. The solid and dashed vertical lines indicate the current age (6 $\pm$ 1 Ga; \citealt{morel18}) of the $\alpha$ Cen AB system.} 
	}
	\label{fig_heat}
\end{figure}

\textcolor{black}{To some extent, we can \textcolor{black}{explore implications from} the peculiar features of $\alpha$ Cen AB on the distributions of [Eu/Fe] and [Eu/Si] \textcolor{black}{for} radiogenic heat production in putative terrestrial-like planets in the system (i.e. "$\alpha$-Cen-Earth").} 

\textcolor{black}{First of all, due to the common origin of the binary stars and their overlapped [Eu/Si] and [Fe/Si] abundances (shown in Fig. \ref{fig_comp_Si}), we do not distinguish such a putative planet around A or B, but \textcolor{black}{compute} the weighted average of the Eu abundances in A \& B to construct an average $\alpha$-Cen-Earth of the system.} 

\textcolor{black}{Second, we need to turn the Eu abundance to the abundances of the $r$-process radioactive elements Th and U (we will discuss K separately). We adopt the observational average \textcolor{black}{($0.014\pm0.045$ dex)} of [Th/Eu] \textcolor{black}{values} of the solar analogues \textcolor{black}{on the Zero-Age Main Sequence, ZAMS} \citep{botelho19} to \textcolor{black}{at first compute} Th abundance from Eu. Then the nearly constant $^{238}$U/Th production ratio ($0.571^{+0.037}_{-0.031}$) in meteorites and Galactic halo stars \citep{Dauphas2005} is adopted to obtain the abundance of $^{238}$U. Although the isotopic ratio $^{235}$U/$^{238}$U may not be necessarily constant \textcolor{black}{during} the Galactic history, its variance, however, is limited \textcolor{black}{when considering} \textcolor{black}{the time interval between 6 and 9 Ga} of the Galactic evolution \citep{Frank2014}\textcolor{black}{; it was within this interval} when both the solar system and the $\alpha$ Cen system formed. Hence, we \textcolor{black}{prefer to take a simple approach and adopt} the well-known\textcolor{black}{, initial} solar system $^{235}$U/$^{238}$U ratio (24.286/75.712; \citealt{Lodders2009}), \textcolor{black}{set for} 4.56 Ga ago, to infer the abundance of $^{235}$U for $\alpha$ Cen AB. We \textcolor{black}{admit that this is} a crude simplification, and we discuss this caveat further in the subsequent section.} 

\textcolor{black}{Third, since both U and Th are refractory lithophile elements (RLEs), their stellar abundances, upon normalising to a major rock-forming RLE (e.g. Mg or Si), \textcolor{black}{implicitly reflect} their concentrations in the primitive mantle\footnote{\textcolor{black}{Primitive mantle = present-day mantle + crust.}} of a rocky planet. Here, we prefer Mg over Si because of its more lithophile nature than silicon (although Si is a lithophile element, it is also widely recognised as a major light element in the core; \citealt{Wang2018} and references therein). On this basis, we can obtain the mass ratio of each individual radionuclide (X) to Mg (i.e. X/Mg). Independently, we can calculate X/Mg of the (primitive) mantle of the Earth based on the literature. With the known mantle concentrations (by mass) of individual radionuclides in the Earth (see Table \ref{tab_heat_earth}), we can scale them by the relative X/Mg between $\alpha$-Cen(-Earth) and the Earth to get the corresponding nuclide concentrations in the primitive mantle of an $\alpha$-Cen-Earth. For simplicity, the mantle mass of $\alpha$ Cen Earth is assumed to be equal to that of the Earth \textcolor{black}{(this assumption is further discussed in Sect. \ref{sect_disc_limit}.4)}.} 

\textcolor{black}{Furthermore, based on the radioactive decay and their known half-lives \citep{Turcotte2002} as well as the age of the $\alpha$ Cen AB system ($6\pm1$ Ga; M18), we \textcolor{black}{model} the radionuclide concentrations in the mantle of such an $\alpha$-Cen-Earth over the geological time. For \textcolor{black}{the normalisation purpose}, the concentration of the stable and major-rock forming element Mg in the mantle is kept unchanged.} 

\textcolor{black}{Eventually, with the heat generation rates (per unit mass of the individual radionuclides\textcolor{black}{; \citealt{Dye2012})}, we \textcolor{black}{compute} the heat output for both the nascent and present-day $\alpha$-Cen-Earth. We have summarised our model constraints (with references) as well as the concentration and radiogenic heating estimates in Table \ref{tab_heat_alpha}. We have also reported our corresponding calculations for the Earth in Table \ref{tab_heat_earth} for comparison.}

\textcolor{black}{Now, we will discuss $^{40}$K, which is not directly constrained from Eu but instead, is based on a GCE assumption for cosmochemically Earth-like planets \citep{Frank2014}.} 

\textcolor{black}{\textcolor{black}{Potassium} is not only irrelevant to Eu (in terms of abundances) due to its distinct production and destruction pathways \citep{Clayton2003, Zhang2006} but also is a volatile element, which means its abundance in a planet hosting star cannot be directly reflected into a rocky planet (due to devolatilisation during planet formation; \citealt{Wang2018b}\textcolor{black}{, and references therein}). Simultaneously, \textcolor{black}{potassium is} among a few most challenging (rock-forming) elements to be \textcolor{black}{widely/}accurately measured in \textcolor{black}{the photospheres of} stars\textcolor{black}{, owning to the fact that only the strong \ion{K}{i} \mbox {7698.96 \AA{}} line -- that is difficult to model properly and strongly affected by non-LTE effects -- is measurable \citep{Reggiani2019, Takeda2019} and that it is not covered by most spectrographs (e.g. HARPS).} Therefore, \textcolor{black}{at the present time,} we \textcolor{black}{are forced} to make crude assumptions on the abundance of K or more precisely, $^{40}$K. In light of the starting mantle concentrations of $^{40}$K ($C(^{40}$K)) for cosmochemically Earth-like planets (i.e. assuming their volatile depletion scales relative to their host stars \textcolor{black}{to be the} same as \textcolor{black}{that of} the Earth to the Sun) as a function of the time after galaxy formation \citep{Frank2014}, we note that the variance in $C(^{40}$K) is \textcolor{black}{insignificant} within about 6-9 Ga into the Galactic history. As such, we directly \textcolor{black}{translate} the concentration of $^{40}$K in the primitive mantle of $\alpha$-Cen-Earth with that \textcolor{black}{of} the primitive mantle of the \textcolor{black}{early} Hadean Earth (4.56 Ga ago). For the instantaneous radionuclide concentrations and heat production (due to the radioactive decay), the same processes mentioned earlier will apply (see more details in Table \ref{tab_heat_alpha} and footnotes therein).} 

\textcolor{black}{Figure \ref{fig_heat} illustrates the calculations of both mantle concentrations and heat output of individual radionuclides ($^{232}$Th, $^{235}$U, $^{238}$U, and $^{40}$K) in the putative $\alpha$-Cen-Earth over geological time (\textcolor{black}{logically presumed} equal to the system age). In terms of the mantle concentrations, $^{40}$K is the most abundant \textcolor{black}{heat-producing nuclide} since planet formation until around the age of 4 Ga, when $^{232}$Th \textcolor{black}{becomes dominant}. $^{238}$U is \textcolor{black}{relatively} modest over geological time while $^{235}$U has become negligible since around 1.5-2 Ga. These trends in concentrations are not exactly identical to those in the heat output of these radionuclides, due to their different heat generation rates. \textcolor{black}{Prominently}, $^{235}$U contributed a significant amount of heating (only a factor of 2 lower than the highest contributor -- $^{40}$K) \textcolor{black}{at} planet formation and \textcolor{black}{only} became \textcolor{black}{negligible} since about 4 Ga. $^{238}$U has started to play the most significant role in heat production from around the age of 3.5 Ga, when $^{40}$K \textcolor{black}{underwent its demise} and then was further \textcolor{black}{subordinated} by $^{232}$Th \textcolor{black}{by the time the system reached} 4.5 Ga. Thereafter, $^{238}$U and $^{232}$Th \textcolor{black}{dominate} the radiogenic heat output by contributing approximately equally \textcolor{black}{up to about} 10 Ga -- the maximum time we have modelled, with $^{232}$Th gradually \textcolor{black}{taking over heat production in all old terrestrial-type planets \citep{Frank2014} owning} to its longest half-life (14 Ga).}

\textcolor{black}{In comparison with the total radiogenic heat output of the Earth (see Tables \ref{tab_heat_alpha} and \ref{tab_heat_earth}), the $\alpha$-Cen-Earth generated $73.4^{+8.3}_{-6.9}$ TW \textcolor{black}{at time of} formation and $8.8^{+1.7}_{-1.3}$ TW \textcolor{black}{at} the present day. \textcolor{black}{We note that this is} $23\pm5$\% and $54\pm5$\% lower than that \textcolor{black}{estimated for} the Hadean Earth ($94.9\pm5.5$ TW) and modern Earth ($19.0\pm1.1$ TW). \textcolor{black}{In our gedankenexperiment, $\alpha$-Cen-Earth is intrinsically less geologically active than the Earth, overall.}} 

\textcolor{black}{\subsection{Uncertainty/limitation analysis}}\label{sect_disc_limit}
\textcolor{black}{Our reported uncertainties associated with the radionuclide concentrations and heat output mainly come from the uncertainties on the stellar abundances (Eu and the reference element Mg), those associated with our model constraints, the uncertainties on the mantle concentrations of the Earth, and the uncertainty of the stellar age.} \textcolor{black}{The uncertainties of heat output from the individual radionuclides are added in quadrature to estimate the uncertainty of the total heat output.} 
\textcolor{black}{It is noteworthy that uncertainties in a relative sense are smaller than those in absolute values. For example, the comparison of heat output between $\alpha$-Cen-Earth and the Earth has yielded smaller error bars on their relative differences ($23\pm5$\% and $54\pm5$\%) than those on the absolute heat output values ($73.4^{+8.3}_{-6.9}$ and $8.8^{+1.7}_{-1.3}$) of $\alpha$-Cen-Earth, since in the latter case the uncertainties on the reference (the Earth) have also been included.}

\textcolor{black}{The accuracy of our model \textcolor{black}{rests} on the \textcolor{black}{robustness} of our assumptions. We adopted the constant values from \cite{botelho19} and \cite{Dauphas2005} for [Th/Eu] and [$^{238}$U/Th], respectively. We \textcolor{black}{used} a constant value for [Th/Eu] \textcolor{black}{following} \cite{botelho19}. However, as we noted earlier, the thorium abundance is determined through modelling of a single line that is also heavily blended with other elemental lines, limiting the accuracy of the modelled abundance \textcolor{black}{regardless of the applied treatments}.} 
\textcolor{black}{Similar concerns also apply to the U/Th ratio from \cite{Dauphas2005}, which may be further affected by the representativeness of \textcolor{black}{the} dominant sample of halo stars \textcolor{black}{chosen} (even though meteoritic data have also been considered) for the solar-like systems. \textcolor{black}{We emphasize} that such measurements/modellings on long-lived radionuclides \textcolor{black}{are, to our knowledge, the best available}. 
For $^{235}$U/$^{238}$U, it may be \textcolor{black}{more appropriate} to run GCE corrections on the adopted early solar system value (24.286/75.712; \citealt{Lodders2009}), in spite of our observation of its limited variance for stars like the Sun and $\alpha$ Cen AB that were formed 6-9 Ga into the Galactic history \citep{Frank2014}. In a more recent work on GCE, \cite{Cote2019} reported \textcolor{black}{an unusually large} error bar (60\%) on the \textcolor{black}{initial} solar system $^{235}$U/$^{238}$U ratio. If we incorporate it into our calculations, it would have an impact on our reported \textit{error bars} for the starting concentration and heat output of $^{235}$U by a factor of about 2, which however is not significant to the total heat output over geological time.} 

\textcolor{black}{In addition, since we have adopted the concept "cosmochemically Earth-like planets" \citep{Frank2014} for \textcolor{black}{considering} $^{40}$K, it \textcolor{black}{limits} our conclusions to the total heat budget \textcolor{black}{of} habitable-zone terrestrial-like planets. \textcolor{black}{These} are supposed to receive the equivalent \textcolor{black}{irradiation} from their parent stars as the Earth from the Sun \textcolor{black}{\citep{Hart1979}}. Considering \textcolor{black}{that} planet formation processes may alter the primordial devolatilisation scale controlled by evaporation/condensation out of a nebular gas \textcolor{black}{\citep{Grossman1974a, ONeill2008, Albarede2009, Hin2017, Norris2017, Wang2019a, Fegley2020}}, such planets \textcolor{black}{might be expected to} have shared similar formation histories as our Earth's, at least in terms of the final effect on the volatile depletion. For this reason, we also caution that it is premature to extend our conclusions to the planet(s) orbiting \textcolor{black}{the M5 red dwarf} \object{Proxima Cen} \citep[][]{Anglada2016, damasso20} in view of the still debated origin of this star \citep[e.g.][]{kervella17b,beech17,feng18} and the fact that the transposition of the stellar abundances to planet bulk composition might be dramatically different between Sun-like stars and the chromospherically active, flaring M \textcolor{black}{stars}.} 

\textcolor{black}{Finally, we note that our assumption of the equal mantle mass of a putative $\alpha$-Cen-Earth as that of the Earth is out of convenience only to enable us to focus on the discussion of the heat output dictated by the estimated mantle concentrations of these radionuclides and then make comparisons between the two planets. It is noteworthy that the current detection limits of radial velocity are still far from being resolvable to any Earth-mass, habitable-zone planet: $M \sin{i}$ $\sim$ 50 $M_{\oplus}$ for $\alpha$ Cen A and $\sim$ 8 $M_{\oplus}$ for $\alpha$ Cen B \citep{Zhao2018}. On the other hand, the mantle/core ratios between an $\alpha$-Cen-Earth and the Earth should not be significantly different, in view of the overlapped [Fe/Si] values between $\alpha$ Cen AB and the Sun (Fig. \ref{fig_comp_Si}). However, we will not go further here in assessing how the planetary mass and mantle/core ratio would affect the radiogenic heat budget; instead, we refer the interested readers to \cite{Frank2014} (in particular, their Sect. 5.2 and Fig. 13) for a detailed discussion of such an aspect.}\\

\section{Conclusions}\label{sect_conc}
\textcolor{black}{In the context of studying the potential of Eu as a convenient tool to diagnose exoplanetary radiogenic heat power (in the absence of abundance measurements of long-lived radionuclides in the majority of planet hosting stars), we present a detailed determination of the abundances of this element in the photospheres of $\alpha$ Cen AB -- our nearest \textcolor{black}{Sun-like stars}. Our spectroscopic analysis shows that europium is depleted with respect to iron by $\sim$ 0.1 dex ($\sim$ 25\%) \textcolor{black}{and to silicon by $\sim$ 0.15 dex ($\sim$ 40\%)} compared to solar in both binary components.  A comparison with a large sample of FGK stars shows that such a depletion in $\alpha$ Cen AB appears to be true relative to the majority of these Sun-like stars. This may have important implications to the potential of an $\alpha$-Cen-Earth (a putative terrestrial-like planet in the system) in generating radiogenic heating if our view on using Eu as a proxy for long-lived radionuclides ($^{232}$Th, $^{235}$U and $^{238}$U) is correct. $^{40}$K has to be treated independently due to its distinct nucleosynthesis pathways and its volatile \textcolor{black}{behaviour}.}

\textcolor{black}{We have applied a simple and intuitive approach to quantify the radiogenic heat output propagated from the Eu abundances combined with other assumptions including a GCE model for $^{40}$K. Our first-order estimates \textcolor{black}{lead us to propose} that the radiogenic heat budget in an $\alpha$-Cen-Earth is $73.4^{+8.3}_{-6.9}$ TW upon its formation and $8.8^{+1.7}_{-1.3}$ TW \textcolor{black}{at} the present day, respectively $23\pm5$\% and $54\pm5$\% lower than that in the Hadean Earth ($94.9\pm5.5$ TW) and in the modern Earth ($19.0\pm1.1$ TW). If we assume all other conditions -- especially the primordial gravitational energy (\textcolor{black}{as yet, unconstrained}) -- are not significantly different between the $\alpha$-Cen-Earth and \textcolor{black}{our} Earth, the mantle convection in the $\alpha$-Cen-Earth would be comparably weaker than our \textcolor{black}{planet} over \textcolor{black}{its equivalent evolution history (as the Earth's)}, \textcolor{black}{subduing} its geological activity and by extension, its long-duration habitable potential.}  

\textcolor{black}{\textcolor{black}{The multivariate nature of} planetary evolution is a complex process \citep{Stevenson2004}, but the similar [Fe/Si] ratios between $\alpha$ Cen AB and the Sun (shown in Fig. \ref{fig_comp_Si}) reveals to us \textcolor{black}{at} the first-order that the relative core-to-mantle mass fractions should not be so different between an $\alpha$-Cen-Earth and the Earth. \textcolor{black}{Detailed modelling of bulk compositions and internal structures of such $\alpha$-Cen-Earths will be investigated in our subsequent paper.}} 

\textcolor{black}{In short, we conclude with caution that Eu can be a convenient and \textcolor{black}{practical} tool, along with other constraints, in helping understand the exoplanetary radiogenic heating potential. This \textcolor{black}{may allude} to a population analysis of such an aspect for increasingly \textcolor{black}{discovered rocky exoplanets, of which the host stellar abundances for long-lived radionuclides are \textcolor{black}{as yet seldom} measurable.}} 
 
\begin{acknowledgements}
\textcolor{black}{We thank an anonymous referee for his/her swift and insightful comments. We also thank D. J. Stevenson for useful comments on an earlier version of the manuscript and C. H. Lineweaver for early discussion. This work has been carried out within the framework of the National Centre of Competence in Research PlanetS supported by the Swiss National Science Foundation. H.S.W and S.P.Q acknowledge the financial support of the SNSF. T.M.} acknowledges financial support from Belspo for contract PRODEX PLATO mission development. \textcolor{black}{S.J.M. thanks the Collaborative for Research in Origins (CRiO) at the University of Colorado, which was supported by The John Templeton Foundation
(principal investigator: S. Benner/FfAME): the opinions expressed in this publication
are those of the authors and do not necessarily reflect the views of the John Templeton
Foundation. S.J.M. also acknowledges the NASA Solar System Workings Program, grant
No. 80NSSC17K0732 (principle investigator: O. Abramov/PSI)}. 
This work has made use of the VALD database, operated at Uppsala University, the Institute of Astronomy RAS in Moscow, and the University of Vienna. This research made use of NASA's Astrophysics Data System Bibliographic Services and the SIMBAD database operated at CDS, Strasbourg (France).
\end{acknowledgements}

\bibliographystyle{aa} 
\bibliography{ms_submitted,/Users/seanwhy/Documents/eLibrary/BibsTex/MyPapersBib-aCen}

\clearpage
\onecolumn
\appendix
\textcolor{black}{\section{Radiogenic heat output of the Earth}}
\label{app_heat}

\begin{table*}[!htbp] 
	\textcolor{black}{
		\caption{The estimates of radiogenic heat output in the Earth, respectively upon its formation and present day.}
		\label{tab_heat_earth} 
		\hskip -0.4cm
		\centering
			\begin{tabular}{ccc|c|cc}
				\hline\hline
				& \multicolumn{2}{c|}{Concentration (ppb)} & &\multicolumn{2}{c}{Heat (TW)\tablefootmark{d}}\\ 
				\hline
				Nuclide (X) & \multicolumn{1}{c}{Present day\tablefootmark{a}} & \multicolumn{1}{c|}{Upon formation\tablefootmark{b}} & X/Mg\tablefootmark{c} & \multicolumn{1}{c}{Present day} & \multicolumn{1}{c}{Upon formation} \\
				$^{232}$Th & $74.6\pm6.8$ & $93.5\pm8.5$ & $4.19\pm0.38 \; \times10^{-7}$ & $7.9\pm0.7$ & $9.9\pm0.9$ \\[2pt]
				$^{235}$U &  $0.143\pm0.014$ & $12.7\pm1.3$ & $5.70\pm0.58 \;\times10^{-8}$ & $0.33\pm0.03$  & $29.1\pm2.9$ \\[2pt]
				$^{238}$U & $19.7\pm2.0$ &$39.9\pm4.0$  & $1.79\pm0.18 \;\times10^{-7}$ & $7.5\pm0.8$ & $15.3\pm1.5$ \\[2pt]
				$^{40}$K &$28.2\pm3.0$ & $353.6\pm37.3$ & {$1.59\pm0.17 \; \times10^{-6}$} &$3.2\pm0.3$ & $40.6\pm4.3$ \\[2pt]
				\textbf{Th+U+K} & \bm {$122.6\pm7.6$} & \bm {$499.6\pm38.5$} & \textcolor{black}{...} & \bm {$19.0\pm1.1$} & \bm{$94.9\pm5.5$} \\		
				\bottomrule 
			\end{tabular}
		\tablefoot{
			\tablefoottext{a}{Refer to the mantle concentrations of \cite{Wang2018} (column 3 of \textcolor{black}{their} Table 1), based on which the radioactive elemental concentrations are converted to the radionuclide concentrations by adopting $^{232}$Th/Th=1, $^{235}$U/U=0.0072, $^{238}$U/U=0.9928, and $^{40}$K/K=$1.19\times10^{-4}$ \citep{Turcotte2002} for the present-day bulk silicate Earth.}\\			
			\tablefoottext{b}{$C_0=C_t\times\mathrm{e}^{t\ln2/T_{1/2}}$, where, $C_0$ and $C_t$ represent the abundances of a radionuclide upon the Earth formation ($t=4.56$ Ga) and present day, respectively; $T_{1/2}$ \textcolor{black}{is given in} Table \ref{tab_heat_alpha}.}\\
			\tablefoottext{c}{X/Mg = $C_0$(X)/$C_0$(Mg), which refers to the \textcolor{black}{early} stage of Hadean Earth (i.e. upon the planet formation); $C_0$(Mg) = $C$(Mg) = $22.3\pm0.2$ \% \citep{Wang2018}, i.e., the concentration of Mg \textcolor{black}{is} kept unchanged over gelogical time.}\\ 
			\tablefoottext{d}{Calculated by multiplying the corresponding concentrations (upon planet formation and present day) with the heat generation rates, $h$, of individual radionuclides (see Table \ref{tab_heat_alpha}) .}
		}
	}
\end{table*}

\end{document}